# A Cloud Computing Capability Model for Large-Scale Semantic Annotation

Oluwasegun Adedugbe[1], Elhadj Benkhelifa[1] and Anoud Bani-Hani[2]

[1] Cloud Computing and Applications Research Lab, Staffordshire University, UK
oluwasegun.adedugbe@research.staffs.ac.uk

[2] College of Technological Innovation, Zayed University, UAE
anoud.bani-hani@zu.ac.ue

**Abstract.** Semantic technologies are designed to facilitate context-awareness for web content, enabling machines to understand and process them. However, this has been faced with several challenges, such as disparate nature of existing solutions and lack of scalability in proportion to web scale. With a holistic perspective to web content semantic annotation, this paper focuses on leveraging cloud computing for these challenges. To achieve this, a set of requirements towards holistic semantic annotation on the web is defined and mapped with cloud computing mechanisms to facilitate them. Technical specification for the requirements is critically reviewed and examined against each of the cloud computing mechanisms, in relation to their technical functionalities. Hence, a mapping is established if the cloud computing mechanism's functionalities proffer a solution for implementation of a requirement's technical specification. The result is a cloud computing capability model for holistic semantic annotation which presents an approach towards delivering large-scale semantic annotation on the web via a cloud platform.

## 1 Introduction

With cloud computing gaining more prominence by the day and its application to different domains such as forensics [1], security [2] and enterprise resource planning [3], it has become imperative to consider its application to the evolvement of the semantic web as well. The semantic web is an emerging web in which documents on the web are annotated with descriptive data, which provides a meaning and defines a context for the documents. With the annotation, computers can understand different concepts such as people, places, organisations, etc. and process them accordingly. For instance, on the semantic web, 'Jaguar' as an animal and as an automobile can be understood and processed as appropriate based on annotation data. Cloud Computing on the other hand, is a computing paradigm which defines "a model for enabling ubiquitous, convenient, on-demand network access to a shared pool of configurable computing resources such as networks, servers, storage, applications, and services that can be rapidly provisioned and released with minimal management effort or service provider interaction" [4]. While several research efforts have been made regard-ing the interaction between cloud computing and semantic technologies, it is vital to classify the different interaction types and their implications. This classification provides a basis for a better understanding of the two technologies and enhance further interaction efforts among them. The modes of interaction between cloud computing and semantic technologies can be categorised into three types as follows.

## 1.1 Cloud-Based

This defines semantic technologies' usage in the cloud, with a basic level of interaction, in which semantic-driven software and applications are run from a cloud platform. Semantic driven in this context implies that the software or application has been developed based on one or more technologies from the semantic web stack. In most cases, these will be on the Software-as-a-Service (SaaS) model. This level of interaction is only basic, as the software or application can easily be migrated to another platform, either cloud or non-cloud.

## 1.2 Semantic Cloud

This defines semantic technologies' usage for the cloud. A cloud platform can be semantic in nature using semantic technologies to enhance the experience for cloud users. This can be in areas such as the discovery of cloud services, configuration metrics for cloud services, etc. by annotating the different entities with metadata, fostering the efficiency of the cloud platform, with respect to service delivery across the different models; IaaS (Infrastructure-as-a-Service), PaaS (Platform-as-a-Service) and SaaS (Software-as-a-service). Currently, domain-specific ontologies exist for providing metadata for cloud entities' description [5]. However, rich-content domain ontologies for cloud services are yet to become popular.

## 1.3 Cloud-Driven

This defines cloud computing usage for semantic technologies. A deep level of interaction is defined here, in which specific, core cloud computing mechanisms are defined and configured to facilitate specific processes within the implementation of annotation for the semantic web. With scalability being one of the major challenges of the semantic web so far [6], the idea is based on the use of these mechanisms to overcome the prevalent challenges with the required processes for the provision of large-scale semantic annotation on the web. This would imply a semantic web that is dependent on a cloud platform for its delivery. Also, with the massive migration of data and applications to the cloud and the convergence of the web and the cloud [7], it is logical to consider a semantic web facilitated (or driven) by a cloud platform. The semantic annotation capability model proposed in this paper is based on the cloud-driven mode of interaction described above and it has been developed based on a specific methodology. Section 2 of this paper reviews existing work in the area of a semantic web facilitated by the cloud and presents the state-of-the-art with a gap analysis. Based on the gaps identified in Section 2, Section 3 presents a summary of specific requirements for large-scale semantic annotation of web documents. Section 4 reviews and analyses cloud computing mechanisms that are very central to delivering large-scale semantic annotation on the web. Furthermore, it analyses the cloud computing mechanisms that are directly responsible for the required scalability with respect to semantic annotation. These are the load balancer, automated scaling listener and cloud usage monitor. In section 5, a model is presented comprising of a mapping of the large-scale semantic annotation requirements with the core cloud computing mechanisms that would facilitate them. The paper is concluded with Section 6 which provides a summary and recommendations for further work in the area.

## 2 Related Work

From the review of publications focusing on the interaction between cloud computing and semantic technologies, it can be observed that the implementation of cloud computing solutions is proposed for web applications across different sectors such as electronic learning, biomedicine, forensics, etc. The authors are generally of the opinion that cloud computing provides appropriate solutions for the enhancement of the semantic web as it applies to different industry sectors. There are suggestions that it can play an instrumental role in enhancing the sematic web because of its numerous characteristics that offer cloud resources on demand to clients and thereby support different business models, with a cloud computing architecture that enables developers to efficiently build and deploy distributed systems [8]; [9]; [10]. This implies that semantic web applications can be developed for public distribution through a cloud platform. Also, with cloud computing's high-performance ability, relatively low cost, and scalability attributes, it is beneficial for driving the semantic web, presenting an effective paradigm for managing, deploying, and offering services using shared infrastructure [11]. Furthermore, several businesses and organisations use cloud computing services to outsource the maintenance of data and this leads to major financial benefits. However, there has been increasing challenges in managing significant quantities of heterogeneous data while ensuring efficient information retrieval. Therefore, experts deploy semantic technologies such as RDF (Resource Description Framework) to offer data in a standardised format so the data can be retrieved easily and understood by both people and machines [12]. Semantic technologies are particularly essential when it comes to the maintenance of data in cloud computing environments. An example is the use of huge RDF graphs for data storage in Hadoop clusters built using affordable commodity class hardware. The Hadoop's MapReduce framework is efficient and scalable; enhancing its use for handling vast amounts of RDF data in the cloud as compared to other traditional approaches.

The huge increase in the amount of structured data present on the web in semantic formats in recent years has been made possible by two factors. Firstly, the presence of Linked Data which has led to millions of entries description that are based on RDF in data sets, such as UniProt, DBpedia, and GeoNames. Secondly, the idea of data portability has increasingly been embraced by the Web 2.0 community, with the initial efforts producing hundreds of millions of RDF-equivalent data. Cloud computing has enhanced the semantic web by enabling semantic technologies to gain popularity in supporting different business models. The enhancement of software abstractions tends to minimise the complexities associated with hardware architectures to the programmer. Some of these software layers include Hadoop, Pig, and HBase and they have the capacity to implement data structures that are understood by several programmers [13].

The work of [8] stated that cloud computing platforms enable vendors offer feasible solutions for facilitating the sematic web. The scholars presented a mobile cloud infrastructure that can enhance the extraction of semantic data from speech recognition. Furthermore, they presented both an architecture and a real prototype that has been developed on different platforms. They proposed a mobile cloud support for semantic-enriched speech recognition in social care (MoSSCa), which is a semantic-enriched speech recognition infrastructure. It also takes advantage of mobile cloud computing and the semantic web at the Platform as a Service (PaaS) level. Similarly, [10] argue that semantic web technologies have played a significant role over the recent years in several scientific disciplines, including bioinformatics and life sciences that heavily rely on computational infrastructures for the management of large-scale data. According to the authors, the huge acceptance of these technologies is increasingly attributed to their appropriateness in enhancing knowledge management. However, despite the huge acceptance in bioinformatics, numerous concerns have emerged relating to their efficiency in facilitating collaborative environmental requirements. This also applies to scenarios where the research areas are required to create and share knowledge with other areas of biomedicine. Therefore, these scholars proposed a framework known as Collaborative Workspaces in Biomedicine (COWB) to enhance the management of collaborative knowledge in biomedical communities. It is argued that COWB can be deployed on a cloud platform to enhance performance and scalability in sharing biological knowledge.

In the field of e-learning, semantic technologies have been implemented to foster adaptive and personalised learning systems to users. The research by [9] proposed the use of a cloud computing infrastructure as the basic back-end system to maintain ontologies, security, data repositories as well as other required server resources for the delivery of semantic e-learning applications. Also, this is to ensure a synchronous data exchange and update between the different layers of such applications. Lastly, the S4 semantic suite developed by [14] is a cloud platform that delivers a series of semantic-related features as services for data and applications across diverse business sectors. Such features include text analytics, data prototyping, RDF database hosting, linked data management, metadata management, etc. While these are semantic-based and features that can leverage cloud for the semantic web, it focuses on text analytics using natural processing language operations. The delivery of large-scale semantic annotation for web documents which is the mainstay of the semantic web remains a challenge. However, a system that optimises cloud computing for the delivery is believed to be the required technical solution.

## 3 Requirements for Large-Scale Semantic Annotation

The research by [15] defined a document-centric knowledge model for the automatic semantic annotation of web documents on the web. The model demonstrated the impact of annotation in knowledge management for web documents and involved the formulation of seven requirements for the process. Based on that work and others within the domain, [16] defined a set of twelve requirements for large-scale semantic annotation on the web. The requirements are based on the delivery of a cloud-driven, holistic semantic annotation platform to drive the semantic web. The implementation of these requirements within a semantic annotation tool is believed to be vital for the realisation of the semantic web. These large-scale semantic annotation requirements are summarised in Table 1.

Table 1. Requirements for large-scale semantic annotation.

| | REQUIREMENTS | DESCRIPTION |
|---|---|---|
| Preparatory Phase | Concept Extraction | The analysis of text from numerous sources such as data dictionaries, vocabularies, thesaurus, etc. to identify and extract concepts such as people, places, organisations, date, time, etc. |
| | Ontology Population | The storage of ontologies in an ontology server, where they can be dynamically populated with newer concepts from online data sources on demand. |
| | Ontology Selection | The selection of ontologies from an ontology server based on factors such as their domain, scope and profile. Selection could be for ontology engineering processes such as ontology mapping or ontology matching. |
| | Ontology Mapping | The alignment of concepts and entities within several ontologies together to provide a broader meaning and an increased scope for the representation of knowledge in a specific domain. |
| | Annotation Data Storage | The storage of annotation data after generation. Annotation data defines context for actual data and can be stored with the data it annotates or away from it. Common storage formats include RDF (Resource Description Framework) and JSON (JavaScript Object Notation). |
| Annotation Phase | Annotation Data Reuse | Once annotation data is generated and stored for a specific web resource, it can be reused on demand anytime there is a request for the web resource. |
| | Annotation Data Sharing | Web resources with the same data can share annotation data, thereby reducing processing time and optimising computing resources such as storage, CPU, RAM, etc. |
| | Annotation On-the-Fly | The online, real-time annotation of web resources with their annotation data once there's a client request to a web server for the web resource. |
| | | |
| Maintenance Phase | Annotation Data Auto-Update | Consistency is necessary between actual data and its annotation data. Annotation data needs to be automatically updatable when the actual data changes. |
| | Ontology Auto-Update | Ontologies evolve over time and there is the need for an automatic updating mechanism to ensure that annotation data remains content-rich and up to date. |
| | Annotation Data Optimisation | The optimisation of annotation data based on its format and supporting standards. Also, this is necessary to remove semantic redundancies and ambiguities from the data. |
| | Annotation Data Colocation | The storage of application nodes next to each other improves application performance. Compute, storage and application nodes can be collocated next to each other as this minimises network latency. |

## 4  Cloud Computing Mechanisms

Several cloud services run together to bring about cloud infrastructure mechanisms. The fundamental building blocks of the cloud environment, herein called cloud mechanisms, forms primary artefacts that, in turn, forms the fundamental cloud technology architecture. Having reviewed and analysed the requirements for large-scale semantic annotation on the web, we consider how different cloud computing mechanisms can foster their implementation, focusing on mechanisms that directly impact on the delivery of applications via the cloud on a large-scale. These mechanisms facilitate the design of various cloud applications that are reliable, scalable and secure in nature, ensuring that cloud consumers can trust the services offered by cloud providers. In general, cloud computing uses different approaches to achieve the same cloud services. The concept behind cloud computing is to use several servers that are hosted remotely on the internet to store, manage and process data rather than using local servers. Presence of cloud mechanisms does not only standardize proven practices and solutions in a design pattern format, it further adds standardization to pattern application options. As [6] explains, scalability is one of the major issues and aspects that affect cloud operations. They define scalability as the ability of a system to appropriately and effectively fit a specified problem as the scope of the problem increases. In this case, the scope is described with regards to the number of objects or elements, increasing volume of work or being susceptible to enlargement. Dynamically scalable architecture is primarily based on a system of pre-defined scaling condition that triggers the automatic allocation of resources from predefined resource pools. This dynamic and automatic allocation allows and enables for the variable use and utilization of the cloud resources as dictated to by the fluctuations of the usage demand [17]. The scaling listener, whose primary role and function is to assess when there exists the need for scalability, is configured to dictate when a new IT resource is required based on the workload threshold. There exist several types of dynamic scalability architecture. In dynamic horizontal scaling, the instances of IT resources are scaled out or in in order to handle the fluctuating workload. In dynamic vertical scaling, the IT resources are often scaled either up or down in response to a need for ad-justment in the processing capacity of an IT resource [18]. The third type is called dynamic relocation. In this approach to scaling, the IT resources in consideration are re-located to a specified host with more capacity as it requires. In the following sections, three cloud computing mechanisms very crucial for dynamic scalability are analysed. These are load balancing, automated scaling listener and cloud usage monitor.

### 4.1 Load Balancing

The load balancer mechanism is an agent that works on runtime and programmed with basic logic aimed at the employment of horizontal scaling to ensure the neutrality of a workload across two or more IT resources. The balanced workload will foster increased performance and capacity that supersede what is obtainable with a single IT resource. This is achieved by algorithms that divide roles. However, load balancer mechanism can help in the distribution of annotated data on-demand through its other functions. Load balancer performs Asymmetric Distribution, Workload Prioritization, and Content-Aware Distribution [19]. In Asymmetric Distribution, it assigns larger workloads to IT resources with higher processing capacities. Workloads are placed in different priority levels and scheduled, placed in queues, discarded, and distributed accordingly in Workload Prioritization. For Content-Aware Distribution, "requests are distributed to different IT resources as dictated by the request content." The functions performed by a load balancer are due to its location on the communication path between the IT resources generating the workload and the IT resources performing the workload processing [20].

According to [21], load balancing in cloud computing helps with the re-assignment of the whole load to every node within the shared system for building the efficiency of resource utilization. This is also to obtain good response time of the job. Going along is the concurrent removal of a state in which a few of the nodes are overloaded with respect to some of the nodes that are under-loaded. As described by [22], consideration is not given to the early performance of the system by a dynamic load balancing algorithm but is influenced by the current behaviour of the system. However, the key considerations to be given priority when trying to deploy such algorithm are: "evaluation and comparison of load, stability of different system, performance of system, communication between the nodes, nature of job to be transferred and selection of nodes." It is possible to consider the load based on Network loads, CPU load, and amount of memory utilize.

### 4.2 Automated Scaling Listener

The automated scaling listener mechanism is a service agent that helps in monitoring and tracking communications between users and cloud services being accessed in order to achieve dynamic scaling. On a general note, automated scaling listeners are mostly located near the firewall where they engage in automatic tracking of workload status information. The determination of workloads is possible by the volume of "cloud consumer-generated requests or via back-end processing demands" initiated by certain types of requests [19]. Therefore, the major aim of using an auto-scaling mechanism is to enhance the automatic adjustment of acquired resources to minimise cost while satisfying Service Level Agreements [23].

With the aid of this mechanism, it is possible for users to define triggers by specifying performance metrics and thresholds. A user can set thresholds such that when the observed performance metric exceeds or falls below the threshold, there is an addition or subtraction of predefined number of instances from the application. This type of automation helps to enhance the merits of cloud dynamic scalability [24]. It gives the opportunity to have the addition of more resources for the handling of increasing workload and shuts down unnecessary machines to save cost. Hence, there is no need to be worried about capacity planning, as the extent of the available resource can be adaptive to the application real-time workload [25]. Some of the performance metrics used when cloud auto-scaling mechanisms are deployed includes CPU utilisation, disk operation, bandwidth usage, and so on. These metrics are based on the performance of infrastructures and they help in indicating system utilisation information.

### 4.3   Cloud Usage Monitor

The collection and processing of data based on computer resources utilisation is the role of the cloud usage monitor. It comes in different formats and there are three common "agent-based implementation formats" with each having the ability to be assigned the role of forwarding the user data that is collected to a log database to post-process and report in real-time [26]. The three agents that aid implementation formats to make real-time report possible in the cloud usage monitor mechanism are: monitoring agent, resource agent, and polling agent. A monitoring agent is a program that is based on events and an intermediary service agent. It exists along the paths of communication in order to provide transparent monitoring and analysis of the flow of data. As for the resource agent, it helps in the collection of usage data through event-driven communications with certain resource software that is specialized. A polling agent is a processing module that is known for the collection of cloud service usage data by polling IT resources. Polling agents are mostly used in cloud service monitor for a periodic monitoring of IT resource status like uptime and downtime [18].

The many benefits of using cloud usage monitor mechanism apply to both cloud provider and clients. As described in the study of [27], one of the important features of cloud computing services is the ability of cloud providers to have resources monitoring mechanism to monitor the current status of allocated resources. This helps them to provide efficient services to their users by handling future requests and observe malicious usage. This mechanism is also useful to users in terms of analysing their resources requirements as well as being able to obtain value for the cost incurred in cloud resources usage. Further, it gives them the opportunity to have knowledge of the time to request more resources, relinquish resources that are not in use or of little use, and the quantity of various physical resources required and appropriate for the variety of applications they are running.

### 4.4 Classification of Cloud Computing Mechanisms

The work by [19] proposed a classification for cloud computing mechanisms based on their characteristics and technical solutions they provide. This research effort has adopted the classification for providing necessary details regarding roles and responsibilities for the cloud computing mechanisms mapped with large-scale semantic annotation requirements for the semantic web. The classification helps define a scope and context for each of the cloud computing mechanisms. The first classification is for Cloud Infrastructure Mechanisms as being vital to a cloud environment by enabling cloud technology architecture. Considering a cloud environment as an infrastructure, these mechanisms constitute IT solutions on which the infrastructure is built. Examples of these include Failover System, Resource Cluster, Resource Replication and Geotag.

While several cloud computing mechanisms are inherited from traditional computing models or other computing paradigms such as grid and utility computing, some others are native to the cloud computing paradigm and were standardised as technical solutions to specific IT challenges for the cloud environment. Examples of these include Cloud Usage Monitor, Cloud Workload Monitor, Automated Scaling Listener and Load Balancer. There are also cloud management mechanisms, which focus on management tasks required within a cloud environment. Such tasks include the set-up, configuration, maintenance and monitoring of IT systems. Examples include Billing Management System, Resource Management System and SLA Management System. Cloud Security Mechanisms on the other hand are concerned with implementation of security features for a cloud environment to counter security threats. Examples are Digital Certification, Digital Signature, Identity and Access Management and Access Control.

## 5 Semantic Annotation Cloud Capability Model

Having identified the requirements for large-scale semantic annotation for the semantic web (section 3) and cloud computing mechanisms required for their implementation (section 4), we provide a mapping of the requirements with the core cloud computing mechanisms for each one of them. These mappings are based on the technical processes for implementing each of the requirements and the technical functionalities of each of the cloud computing mechanisms used. In each case, the technical process for the requirement is reviewed and examined against each of the cloud computing mechanisms in relation to their technical functionalities. Hence, we provide a mapping if the cloud mechanism's technical functionalities proffer a solution for the implementation of a requirement's technical process. The mapping table is presented in Table 2.

Table 2. Mapping cloud computing mechanisms with large-scale semantic annotation requirements.

| | Concept Extraction | Annotation On-the-fly | Annotation Data Auto-Update | Annotation Data Sharing | Annotation Data Reuse | Annotation Data Optimisation | Annotation Data Storage | Ontology Auto-Update | Annotation Data Co-Location | Ontology Mapping | Ontology Population | Ontology Selection |
|---|---|---|---|---|---|---|---|---|---|---|---|---|
| Resource Cluster | | ✓ | | | ✓ | | ✓ | | ✓ | | ✓ | |
| Resource Replication | | ✓ | | | ✓ | | ✓ | | ✓ | | ✓ | |
| Failover System | ✓ | ✓ | ✓ | ✓ | ✓ | ✓ | ✓ | ✓ | ✓ | ✓ | ✓ | ✓ |
| Geotag | | ✓ | | | | | ✓ | | ✓ | | ✓ | |
| Automated Scaling Listener | ✓ | ✓ | ✓ | ✓ | ✓ | ✓ | ✓ | ✓ | ✓ | ✓ | ✓ | ✓ |
| Cloud Usage Monitor | ✓ | ✓ | ✓ | | ✓ | ✓ | ✓ | | | ✓ | ✓ | ✓ |
| Load Balancer | ✓ | ✓ | ✓ | ✓ | ✓ | ✓ | ✓ | ✓ | ✓ | | ✓ | |
| Cloud Workload Scheduler | ✓ | ✓ | ✓ | | ✓ | ✓ | ✓ | ✓ | ✓ | ✓ | ✓ | |
| Billing Management System | ✓ | ✓ | ✓ | | | ✓ | ✓ | ✓ | ✓ | ✓ | ✓ | |
| SLA Management System | ✓ | ✓ | | ✓ | | | ✓ | ✓ | ✓ | ✓ | ✓ | ✓ |
| Resource Management System | | ✓ | ✓ | ✓ | ✓ | ✓ | ✓ | ✓ | ✓ | ✓ | ✓ | ✓ |
| Attribute-Based Access Control | ✓ | | ✓ | ✓ | | | ✓ | ✓ | | | ✓ | ✓ |
| Digital Certification | ✓ | ✓ | | ✓ | ✓ | | | | ✓ | | | |
| Digital Signature | ✓ | ✓ | | ✓ | ✓ | | | | ✓ | | | |
| Identity and Access Management | | | | ✓ | ✓ | | ✓ | | ✓ | | ✓ | ✓ |

From the above table, it can be observed that the requirements for large-scale semantic annotation on the web can be facilitated using a cloud architecture which has been specifically designed for this purpose and that considers the specific needs for each of the requirements. These mechanisms presented in the table above are implemented as a core suite for this objective alongside other necessary cloud computing mechanisms for deploying applications on the cloud. Different cloud architectural models emphasize cloud characteristics to varying degrees and deploy cloud design patterns and mechanisms accordingly to meet the requirements of the cloud characteristics [19]. The implication of that is that the patterns and mechanisms can be implemented for increased processing power with compromise on application functionality. However, with a cloud architecture that is specifically designed for large-scale semantic annota-tion on the web, such issues can be avoided. Hence, from the above table, we develop a cloud capability model for large-scale semantic annotation, as follows.

| Ontology Population<br>A (1, 2, 3, 4)   B (1, 2, 3, 4)<br>C (1, 2, 3)   D (1, 4) | Concept Extraction<br>A (3)   B (1, 2, 3, 4)<br>C (1, 2)   D (1, 2, 3) |
|---|---|
| Ontology Auto-Update<br>A (3)   B (1, 2, 3, 4)<br>C (1, 2, 3)   D (1) | Annotation On-the-fly<br>A (1, 2, 3, 4)   B (1, 2, 3, 4)<br>C (1, 2, 3)   D (2, 3) |
| Ontology Mapping<br>A (3)   B (1, 2, 3, 4)<br>C (1, 2, 3)   D (0) | Annotation Data Storage<br>A (1, 2, 3, 4)   B (1, 2, 3, 4)<br>C (1, 2, 3)   D (1, 4) |
| Ontology Selection<br>A (3)   B (1, 2, 3)<br>C (2, 3)   D (1,4) | Annotation Data Auto-Update<br>A (3)   B (1, 2, 3, 4)<br>C (1, 3)   D (1) |
| Annotation Data Optimization<br>A (3)   B (1, 2, 3, 4)<br>C (1, 3)   D (0) | Annotation Data Re-Use<br>A (1, 2, 3)   B (1, 2, 3, 4)<br>C (3)   D (2, 3, 4) |
| Annotation Data Sharing<br>A (3)   B (1, 3)<br>C (2, 3)   D (1, 2, 3, 4) | Annotation Data Co-Location<br>A (1, 2, 3, 4)   B (1, 3, 4)<br>C (1, 2, 3)   D (2, 3, 4) |
| **A. Cloud Infrastructure Mechanisms**<br>1. Resource Cluster<br>2. Resource Replication<br>3. Failover System<br>4. Geotag<br><br>**B. Specialized Cloud Mechanisms**<br>1. Automated Sealing Listener<br>2. Cloud Usage Monitor<br>3. Load Balancer<br>4. Cloud Workload Scheduler | **C. Cloud Management Mechanisms**<br>1. Billing Management Systems<br>2. SLA Management System<br>3. Resource Management System<br><br>**D. Cloud Security Mechanisms**<br>1. Attribute-Based Access Control System<br>2. Digital Certification<br>3. Digital Signature<br>4. Identity and Access Management |

**Fig. 1.** Cloud Capability Model for Semantic Annotation.

The above model is based on the principles of semantic annotation. Semantic annotation is a data integration technique for actual data, such as data from web documents and metadata; from multiple sources in such a way that results in a unified view of the combined data. It consists of twelve functionalities that are required for cloud-driven semantic annotation of web documents. From the above model, it can be observed that cloud computing has a significant role to play in the actualisation of a truly semantic web; one in which documents and resources on the world wide web can be provided with the required rich-content semantic annotation for their data, based on a cloud model designed for the purpose. The semantic annotation of web documents thereby, results in a context-aware web, in which web documents and resources are processed as "things" rather than as "strings".

Each of the requirements in the model can be observed to require cloud computing mechanisms specifically configured for them towards the delivery of a holistic semantic annotation that considers the required associated features needed for an ideal and complete cycle of real-time, continuous and dynamic semantic annotation. While several applications run simply as cloud-based; being easily detachable from the cloud platform on which they are running, the above model considers a semantic annotation application that is dependent on the cloud platform on which it is running. This is to ensure that the design pattern for the cloud architecture has been configured and optimised for the delivery of large-scale semantic annotation on the web, leveraging cloud computing capabilities optimally for meeting the semantic annotation requirements. The cloud computing mechanisms implemented to deliver the design pattern are configured to deliver optimal performance for semantic annotation. Based on these, the above model provides an approach for implementing a cloud architecture that will deliver 'cloud-driven' semantic annotation on a large to documents and resources on the world wide web. Each concept is facilitated by four categories of cloud computing mechanisms. Based on [18] and [19], a summary description for cloud computing mechanisms in Fig. 1 is presented in Table 3.

**Table 3.** Description for the cloud computing mechanisms.

| | Cloud Computing Mechanisms | Description |
|---|---|---|
| Cloud Infrastructure Mechanisms | Resource Cluster | Helps in grouping multiple IT resource instances for operation as a single IT resource. This leads to an increase in the combined computing capacity, load balancing, and availability of the clustered IT resources. |
| | Resource Replication | For creating multiple instances of the same IT resource and is usually done when the availability of an IT resource and performance need to be improved. The implementation of this mechanism to replicate cloud-based IT resources is carried out using virtualization technology. |
| | Failover System | Increases the reliability and availability of IT resources with the aid of established clustering technology for the provision of redundant implementations. It is usually programmed for automatic switch over to a redundant or standby IT resource instance in situation where the currently active IT resource is no longer available. |
| | Geotag | A "data receptacle in a trusted platform module (TPM)." It holds geolocation properties and makes provision for the mechanism with aid of its geolocation feature. It has the capability to tag data on the event of server provision- |
| | | ing the first time in a data centre. |
| **Specialised Cloud Mechanisms** | Automated Scaling Listener | A service agent that helps in monitoring and tracking communications between users and cloud services being accessed to achieve dynamic scaling. On a general note, automated scaling listeners are mostly located near the firewall – where they engage in automatic tracking of workload status information. |
| | Cloud Usage Monitor | An independent self-governing software program whose function is to collect and process data on how IT resource is utilized. It is not complex, and it works based on the kind of usage metrics it is programmed to collect as well as the method of data collection. |
| | Load Balancer | An agent that works on runtime and programmed with basic logic aimed at the employment of horizontal scaling to ensure the neutrality of a workload across two or more IT resources. The balanced workload will foster increased performance and capacity that supersede what is obtainable with a single IT resource. |
| | Cloud Workload Scheduler | For automating, monitoring, and controlling the workflow within the cloud infrastructure. The automating feature of the mechanism is used for the management of vast number of workloads per day from a single point of control. |
| **Cloud Management Mechanisms** | Billing Management System | For collecting and processing usage data pertaining to cloud provider accounting and cloud consumer billing. Basically, the billing management system works with pay-per-use monitors for gathering runtime usage data that is stored in a repository that the system components then draw from for billing reporting and invoicing purposes. |
| | SLA Management System | Aids specific observation of runtime performance of services offered in cloud environment for ensuring the fulfilment of what is required in the contractual QoS as stated in SLAs. |
| | Resource Management System | Helps in the coordination of IT resources based on management actions performed by both cloud consumers and cloud providers |
| **Cloud Security Mechanisms** | Attribute-Based Access Control | An access control method that guarantees the allowance or denial of access to resources in the cloud by consumers on demand or request "based on attributes of the consumer, attributes of the resource, environment conditions, and a set of policies that are specified in terms of those attributes and conditions." |
| | Digital Certification | It makes use of what is known as certificate. "A certificate is a data file that binds the identity of an entity to a public key and contains the user's identification and a signature from the issuing authority |
| | Digital Signature | Helps with the provision of data authenticity and integrity using authentication and non-repudiation. A digital signature makes provision for the evidence that the message received is not different from that created by its rightful sender. |
| | Identity and Access Man- | Helps in controlling and tracking the identity of users and |
| | agement | access privileges for IT resources, environments, and systems. |

## 6   Conclusions and Future Work

This paper discussed different cloud mechanisms and how they can impact on a cloud-driven semantic annotation solution. Based on these, we have proposed a cloud semantic annotation capability model that provides a mapping of large-scale semantic annotation requirements with core cloud computing mechanisms that will facilitate them. It is believed that the adoption of this model for the semantic web in a cloud computing environment will provide a means of delivering semantic annotation for web documents as a cloud service based on a semantic annotation technique. However, further research is required to provide a finer granularity for the interactions between the requirements and the different cloud computing mechanisms, in which sub-mappings can be identified and a functional specification designed for their implementation. Finally, it is expected that the cloud semantic annotation capability model will generate further discussions on the development and deployment of cloud capability models for other concepts in computing.